\documentclass[twocolumn, iop,aps,superscriptaddress,showpacs,floatfix]{revtex4-1}

\usepackage{mathptmx}
\usepackage{subfigure}
\usepackage{dcolumn}
\usepackage{amsmath,amssymb}
\usepackage{bm}
\usepackage{color}
\usepackage{lineno}

\usepackage{latexsym}
\usepackage{epstopdf}
\usepackage{color}
\usepackage{mdframed}
\usepackage[english]{babel}
\usepackage{latexsym}

\usepackage{psfrag,graphicx}
\usepackage{epsf}
\usepackage{subfigure}
\usepackage{amsfonts}
\usepackage{bm}
\usepackage{natbib}
\usepackage{epstopdf}
\usepackage{appendix}
\usepackage{changes}
\usepackage{bbm}

\definecolor{mygrey}{gray}{0.35}
\definecolor{myblue}{rgb}{0.2,0.2,0.8}
\definecolor{myzard}{cmyk}{0,0,0.05,0}
\definecolor{mywhite}{rgb}{1,1,1}
\definecolor{myred}{rgb}{1,0.,0.3}

\usepackage[colorlinks=true,citecolor=myblue,linkcolor=myred]{hyperref}

\def\be{\begin{equation}}
\def\ee{\end{equation}}
\def\ba{\begin{align}}
\def\enda{\end{align}}
\def\bi{\begin{itemize}}
\def\ei{\end{itemize}}

 \def\ee{\mathord{\rm e}}

 \def\ee{\mathord{\rm e}}

\renewcommand{\ee}{{\rm e}}

\def\beq{\begin{equation}}
\def\beq{\begin{equation}}
\def\eeq{\end{equation}}

 \newcommand{\ket}[1]{|#1\rangle}
 \newcommand{\bra}[1]{\langle #1|}

\begin{document}

\title[Short Title]{On the robustness of  the NV - NMR spectrometer setup to magnetic field inhomogeneities}
\author{Yotam Vaknin}
\affiliation{Racah Institute of Physics, The Hebrew University of Jerusalem, Jerusalem
91904, Givat Ram, Israel}
\author{Benedikt Tratzmiller}
\affiliation{Institut fur Theoretische Physik und IQST, Albert-Einstein-Allee 11, Universitaet Ulm, D-89081 Ulm, Germany}
\author{Tuvia Gefen}
\affiliation{Racah Institute of Physics, The Hebrew University of Jerusalem, Jerusalem
91904, Givat Ram, Israel}
\author{Ilai Schwartz}
\affiliation{Institut fur Theoretische Physik und IQST, Albert-Einstein-Allee 11, Universitaet Ulm, D-89081 Ulm, Germany}
\author{Martin Plenio}
\affiliation{Institut fur Theoretische Physik und IQST, Albert-Einstein-Allee 11, Universitaet Ulm, D-89081 Ulm, Germany}
\author{Alex Retzker}
\affiliation{Racah Institute of Physics, The Hebrew University of Jerusalem, Jerusalem
91904, Givat Ram, Israel}
\date{\today}


\begin{abstract}
The NV-NMR spectrometer is a promising candidate for detection of  NMR signals at the nano scale.
Field inhomogeneities, however, are a major source of noise that limits spectral resolution in state of the art NV - NMR experiments and constitutes a major bottleneck in the development of nano scale NMR.
Here we propose, a route in which this limitation could be circumvented in NV-NMR spectrometer experiments, by utilising the nanometric scale and the quantumness of the detector. 
\end{abstract}
\maketitle

{\em Introduction ---} Nuclear Magnetic Resonance (NMR) spectroscopy can identify the magnetic frequencies associated with specific atoms, bonds or molecules\cite{slichter2013principles} and thus is used ubiquitously for structure and chemical analysis. It can estimate frequencies with very high precision, but needs relatively large sample sizes. Thus, considerable efforts have been invested in decreasing the size of an NMR sensing region.

The simplest way to decrease the minimal sample volume is to decrease the size of the measuring coil\cite{lacey1999high,grisi2017nmr,glover1994microscope}. These micro-coil setups are susceptible to noise in the frequency domain due to field inhomogeneities and variations in magnetic susceptibilities. These issues result in line broadening of atomic spectra, that limits the NMR precision.

Another way to decrease the minimal size of the probe is to use the NV - NMR spectrometer\cite{schmitt2017submillihertz,boss2017quantum,glenn2018high,aslam2018nanoscale,degen2009nanoscale,staudacher2013nuclear,mamin2013nanoscale,devience2015nanoscale,staudacher2015probing,laraoui2013high,muller2014nuclear} which is based on the nitrogen-vacancy (NV) quantum defect in diamond.
This promising direction enables reductions in the sample volumes by several orders of magnitude. Because of the local nature of the detection of each NV, one might expect a significant sensitivity to field inhomogeneities and local fluctuations as is indeed the case in \cite{glenn2018high,schmitt2017submillihertz,boss2017quantum}. We will show here that with suitable detection protocols, the NV - NMR spectrometer can be made extremely robust to field inhomogeneities, thus, overcoming the main bottleneck in micro/nano-NMR setups.

\begin{figure}[bp]
\includegraphics[height=4.7cm, width=7.5cm]{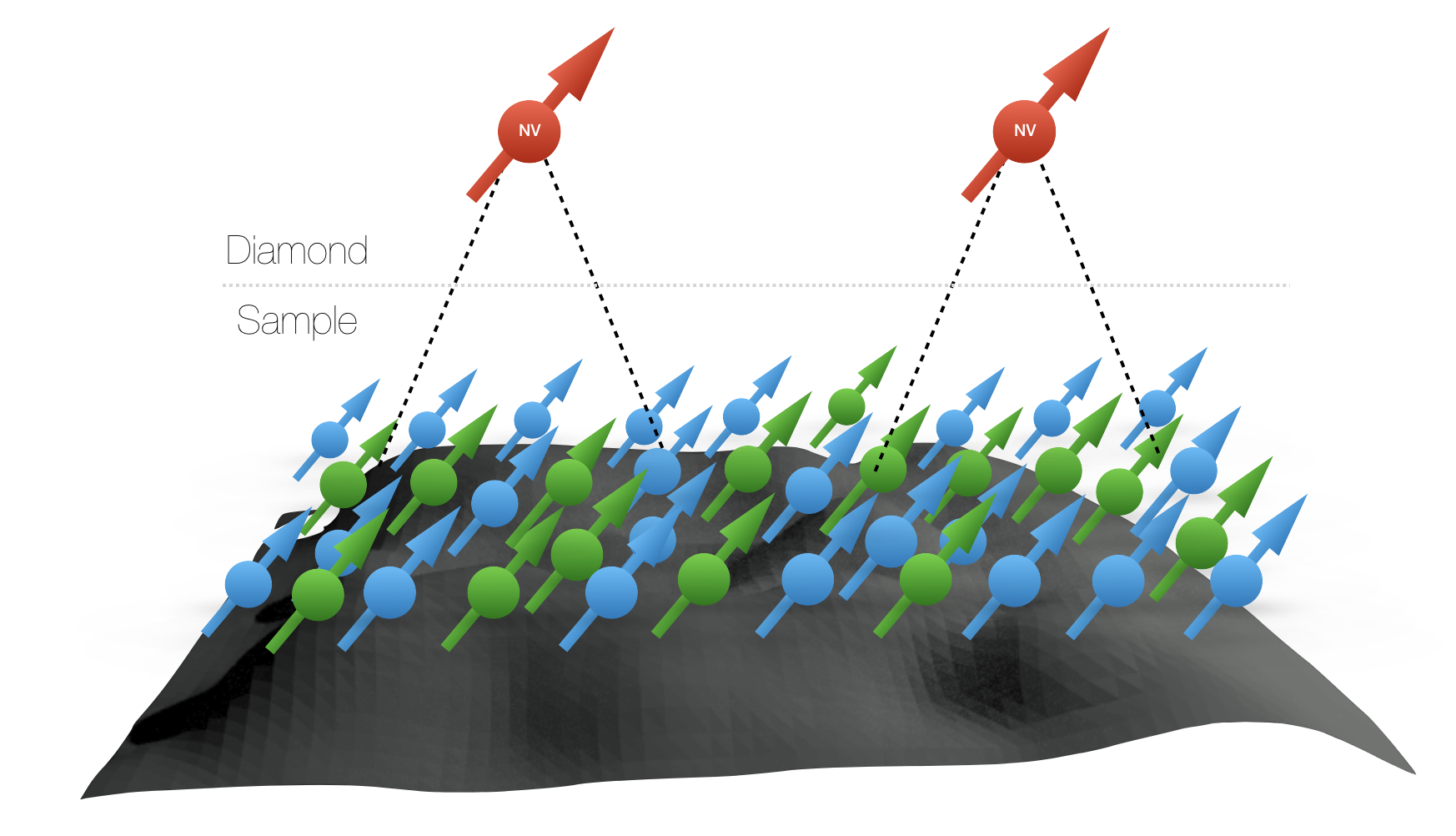}
\caption{NV Probe}
\label{figure:NVProbe}
\raggedright
A diamond doped with NV centers measures the magnetic field created by two populations of atomic dipoles in a sample. The two populations of dipoles, labeled by blue and green, each rotate with a different frequency. The black surface represents a projection of the magnetic field, to show its inhomogeneities, which affect each location in the sample differently and change the dipole's rotation frequency.
The dotted lines indicate the region sampled by a single NV. The small magnetic fluctuation in this region will affect the resolution. The large fluctuations between the regions, however, will be canceled out by the method presented here. 
\end{figure}

Any type of nanometric probe will solve the inhomogeneity problem, since it will measure a small spatial region in which there are only slight variations in the field, as shown in fig.\ref{figure:NVProbe}.
Whenever creating a grid of such probes, with no ability for individual readout, the final signal will again average noise and inhomogeneities over the entire region, resulting in the same problem.
We will show that quantum probes, as a result of their inherent non linearity,  do not suffer from this problem and will almost exclusively only be affected by the tiny gradients that smear the signal in the region read by a single probe.
This noise can reasonably be assumed to be smaller by orders of magnitude than the total noise on the whole sample.

In this paper we demonstrate how such a scheme can be carried out, by only changing the measurement base of current NV based NMR experiments, thus avoiding any technological or experimental overhead. This result can be generalized to hypothetical grids of either conducting coils or NVs that independently measure the signal from each nanometric probe. 

{\em Definition of the problem ---}
We address the main scenario of the NV - NMR spectrometer; i.e., the resolution of very close frequencies which occur, for example, due to chemical shifts or J - couplings of a sample at the nano scale.
As a proof of principle analysis we address the resolution capability of two adjacent frequencies.

The NV magnetic dipole is coupled to a mesoscopic sample of nuclei, that can be modelled semi-classically: To the leading order, the effect of the nuclei on the NV can be modelled by the influence of an oscillating magnetic field. Thus the Hamiltonian that captures the central components of the problem is:
\begin{equation}
H_C = g \sigma_z \big( \sin(\delta_1 t )  +   \sin(\delta_2 t )  \big ),
\label{Ham_Quant}
\end{equation}
where $\delta_1,\delta_2$ are the two frequencies of the sample, up to an offset, $g$ is the effective coupling and $\sigma_z$ is a Pauli matrix.

Before analyzing this Hamiltonian, let us start by introducing the problem that appears in the regular NMR setting. In this case the signal, which is the current,  is proportional to the derivative of the magnetic flux; i.e.,
\begin{eqnarray}
I(t) \propto &g& \delta_1 \cos(\delta_1 t )  + g\delta_2   \cos(\delta_2 t ) \nonumber \\
= &g& \delta_1( \cos(\delta_1 t ) +    \cos(\delta_2 t )   )  + g (\delta_2 - \delta_1)   \cos(\delta_2 t ),
\label{Class_Signal}
\end{eqnarray}
because of the large $\delta_1$ factor, which is due to the high magnetic field, the information on the difference between the frequencies is in the first part; hence, we will concentrate on it.
Due to field inhomogeneities, both signals which originate from the same microscopic region have a random shift ($\epsilon$) in their frequency, but as this shift is common, the signal is proportional to:
\begin{eqnarray}
\label{classical_signal}
&& g \delta_1 \left( \cos(\left(\delta_1+\epsilon \right) t )  +   \cos(\left( \delta_2+\epsilon \right) t )  \right ) \nonumber\\ &&= 2 g \delta_1  \cos\left ( \frac{\delta_1 +\delta_2 + 2\epsilon}{2}  t \right )     \cos \left ( \frac{\delta_1 - \delta_2}{2} t \right ),
\end{eqnarray}
and thus the frequency of the beat note does not change but the carrier frequency is shifted by $\epsilon$.
Since in each microscopic region the shift is different, the full signal (detected by the coil) is obtained by integrating over $\epsilon,$ 
namely integrating over all the microscopic shifts.
Assuming $\epsilon$ is uniformly distributed in a range $\pm \sigma$ yields:
\begin{equation}
I\left(t\right)\propto2g\delta_{1}\cos\left(\frac{\delta_{1}+\delta_{2}}{2}t\right)\cos\left(\frac{\delta_{1}-\delta_{2}}{2}t\right)\frac{\sin\left(\sigma t\right)}{\sigma t}.
\end{equation}
Hence the inhomogeneity imposes a decay time of $1/\sigma$ (set by the range of inhomogeneity).   

The inhomogeneity range is determined by the type of conducting coil. 
 For a single micro-coil, the distribution of the shift is very narrow, on the scale of the microscopic region it probes.
 Denoting the standard deviation of this distribution as $\Delta$, the corresponding decay time scales as $\Delta^{-1}.$
Hence a micro-coil gives rise to a longer decay time (compared to a macroscopic coil); however, due to the smaller integration region, the signal is much weaker.  
Ideally we would like to have the (large) amplitude of a macroscopic coil and the small decay of a microscopic coil. Is this achievable? Here, we show that by employing a grid of NV centers we can achieve a desirable small decay, that scales as the microscopic region probed by a single NV.
Remarkably this does not require single site addressing of the NV centers or any complex measurement or data analysis techniques{\footnote {A possible suggestion would be to use a grid of micro-coils.    
 A grid of individually addressed micro-coils will be more sensitive
and will only suffer from a negligible effect of noise; i.e., $\Delta t.$
However, in the case where each coil cannot be read individually and the only measurable quantity is the sum, the signal will be averaged over the wide ($\sigma$) distribution and thus considerably diminished. }}
.

{\em Quantum case ---}
Let us derive the signal obtained from a grid of NV centers.
Consider first a single NV center: we read the probability of measuring $\ket{\uparrow_x}$ or $\ket{\downarrow_x}$ after initiating the NV in $\ket{\uparrow_x} = \frac{\ket{\uparrow_z} + \ket{\downarrow_z}}{\sqrt{2}}$. The evolution of this state is:
$ \ket{\psi(t)} = \frac{\ket{\uparrow_z} e^{i\phi(t)} + \ket{\downarrow_z} e^{-i\phi(t)}}{\sqrt{2}}, $
where the phase is $\phi(t) = \int_{t-\frac{\tau}{2}}^{t+\frac{\tau}{2}} \mathrm{d}t g(\sin(\delta_1 t) + \sin(\delta_2 t))$, integrated over the length of the interaction ($\tau$).
In this setup, the probability of measuring $\vert \downarrow_{\phi_m} \rangle$ is:

\begin{eqnarray}
P_{\downarrow_{ \phi_{m}  }}&&=\sin^{2}\left( \phi + \frac{\phi_m}{2} \right),
\label{prob1_with_phi}
\end{eqnarray}
where $\phi_m$ denotes the measurement angle (in the Bloch sphere) with respect to x.
Given a microscopic noise ($\epsilon$), the accumulated phase ($\phi$) reads: 
\begin{equation}
\phi = g\int_{t-\frac{\tau}{2}}^{t+\frac{\tau}{2}} \mathrm{d} t\left\langle \sin\left(\left(\delta_{1}+\epsilon\right)t\right)+\sin\left(\left(\delta_{2}+\epsilon\right)t\right)\right\rangle _{\text{microscopic}},
\end{equation}
 where the brackets indicate a microscopic average. We assume the noise ($\epsilon$) is uniformly distributed in a range $\pm\Delta$ around some value $\epsilon_0$. For each frequency this averaging gives:
 \begin{equation}
 \label{microscopic_average}
 \left\langle \sin\left(\left(\delta_{i}+\epsilon \right)t\right)\right\rangle _{\epsilon_0\pm\Delta}=\frac{\sin\left(\Delta t\right)}{\Delta t}\sin\left(\left(\delta_{i}+\epsilon_{0}\right)t\right),
 \end{equation}
which means that the noise inflicts a decay time of $\Delta^{-1}$ and a shift. The expression of the phase, assuming a short interaction ($\tau (\delta + \Delta) \ll \pi$),  gives:

\begin{eqnarray}
\label{Phase}
\phi &&= \left(g\tau\right)\frac{\sin\left(\Delta t\right)}{\Delta t}\left[\sum_{1,2}\sin\left(\left(\delta_{i}+\epsilon_{0}\right)t\right)\right] \\
&&= \left(2g\tau\right)\frac{\sin\left(\Delta t\right)}{\Delta t}\sin\left(\frac{\delta_{1}+\delta_{2}+2\epsilon_{0}}{2}t\right)\cos\left(\frac{\delta_{1}-\delta_{2}}{2}t\right). \nonumber
\end{eqnarray}
From this expression it can be seen that the resolution information is in the beat note, whereas the noise ($\epsilon_0$) is in the carrier frequency. 
In the weak coupling regime, in which $\phi \ll 1$, the probability of measurement when choosing $\phi_m = 0$ is 
{\footnotesize
$
P_{\downarrow_{x}}=\left(2g\tau\frac{\sin\left(\Delta t\right)}{\Delta t}\right)^{2} \sin^{2}\left(\frac{\delta_{1}+\delta_{2}+2\epsilon_{0}}{2}t\right) \cos^{2}\left(\frac{\delta_{1}-\delta_{2}}{2}t\right), 
\label{prob1}
$}which is comparable to the result of a single microscopic coil (Eq. \ref{classical_signal}) averaging only over a $\Delta$ noise distribution. But unlike Eq. \ref{classical_signal},  when averaging the noise ($\epsilon_0$) over the wide noise of the entire sample ($\sigma$), the $\sin^2$ term averages to a half and we get:

\begin{equation}
\label{final_probability}
P_{\downarrow_{x}}=2\left(g\tau\frac{\sin\left(\Delta t\right)}{\Delta t}\right)^{2}\cos^{2}\left(\frac{\delta_{1}-\delta_{2}}{2}t\right).
\end{equation}
In other words, the macroscopic noise does not inflict a decay at all. Note that the short interaction assumption can be easily satisfied in the quantum case, since in this case $\delta$ is not the Larmor frequency but the difference between the Larmor frequency and the control \cite{schmitt2017submillihertz,gefen2017control}.



Eq.\ref{final_probability} implies that an individual readout of the NV's is not necessary.
By measuring the total luminosity, we are in fact averaging $P_{\downarrow_x}$ over the entire sample.
Hence we get a decay of a microscopic sample and the SNR of a macroscopic sample, as desired.
Thus, {\em a diamond based NV nano - NMR setup has the advantages of a microscopic probe even when probing a macroscopic region.}

This analysis describes a situation in which a large ensemble of NV centers close to the surface observe nuclear ensembles that each see a spatially homogenous magnetic field over the detection
volume of a single NV. With magnetic field inhomogeneities existing over the entire
observation region, covered by all NVs. This can be the case for NVs that are implanted some distance
below a planar surface that is smaller than the lateral extent of the NV ensemble. In this situation
a phase insensitive scheme that reads out all NVs will pick up the same signal, resulting from
the frequency difference, which can therefore be added (incoherently). This result is a considerable improvement 
over a Y-readout scheme which would see strongly shifted lines upon averaging 
over many NVs, that would lead to a broadening larger than the chemical shifts.

\begin{figure}
\includegraphics[height=7.5cm, width=7.5cm]{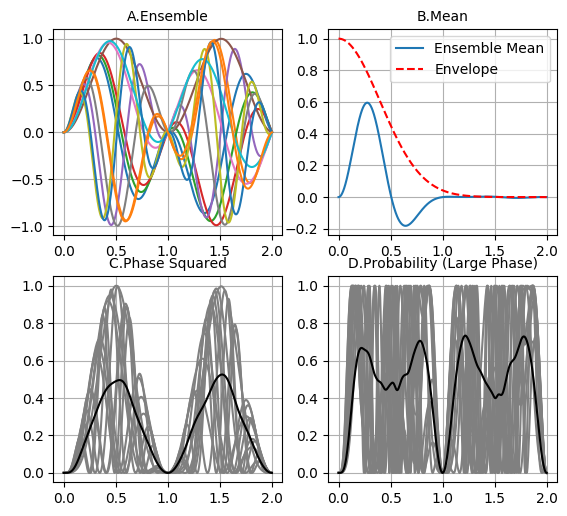}
\caption{Variance Measurement ---}
\label{variance_measurement}
\raggedright
An ensemble of signals that obey Eq \ref{classical_signal}, each with a different $\epsilon$ is shown in (A). The envelope frequency, which is $\delta_1-\delta_2$ can clearly be seen in the ensemble, but it is lost in the decaying average signal $\left\langle \phi \right\rangle$ (seen in (B)) . Since the envelope frequency is visible in (A), its frequency is encoded in the variance of the signal (with respect to the ensemble). A measurement of $\left\langle \phi^2 \right\rangle$ is therefore a measurement of the variance of the signal, which still contains the information about the frequency. Figure (C) shows $\phi^2$ (grey) for different signals in the ensemble, and their mean (black), now with a clear envelope.
In the large phase regime the measured probability is $\sin(\phi)^2$, as can be seen in (D).
This signal contains even more information about the frequency difference (higher harmonics appear as well): we can see a sharp zero in $\pi/\left( \delta_1-\delta_2  \right)$ in the ensemble (grey), and a clear envelope in the average signal (black). In this large phase regime, $\sin(\phi)^2$ can be thought of as the combined  measurements of the various even moments of $\phi$, which all vanish on a beat note of $\delta_1 - \delta_2$. For this reason, the signal is visible independent of the strength of the noise or the coupling. 

\end{figure}

This scheme can be though of as a noise spectroscopy method as in \cite{bar2012suppression,taylor2008high} in which the variance of the phase is measured directly, $\langle \phi^{2} \rangle,$ instead of the average phase.
As can be seen in Fig \ref{variance_measurement}, the information about the frequency difference $\delta_1 - \delta_2$ is not encoded in the averaged phase, but in its variance.
Depending on the measurement basis, we can either measure the averaged phase or its variance, and by measuring the variance the quantum protocol can identify the frequency. In the weak coupling regime ($g\tau \ll 1$), a variance measurement sacrifices some signal since the signal is quadratic in the coupling. To a limited extent, it is possible to regain the signal with an amplifying scheme we will described below.

\begin{figure}[bp]
\includegraphics[height=7.4cm, width=7.4cm]{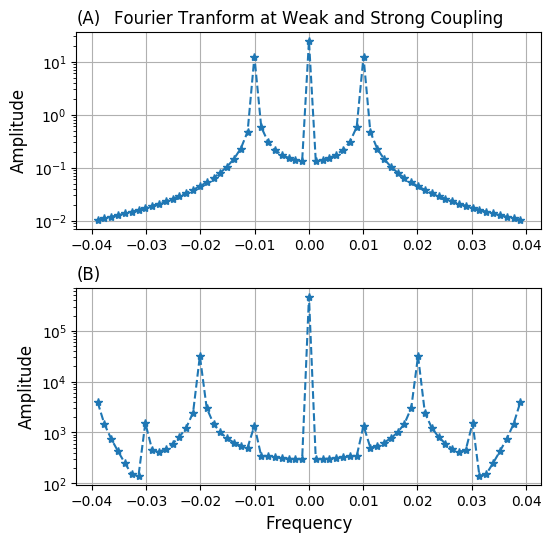}
\caption{Numerical simulation at strong coupling---}
\label{simulation1}
\raggedright
FFT of a numerical simulation of $P_{\downarrow}(t)$ with phase determined by Eq. \ref{Phase} and different couplings $g$, averaged over the entire sample. (A) In the weak coupling regime, it is clear that only a single frequency corresponding to the beat note $\delta_1 - \delta_2 = 10^{-2}$, as expected from Eq. \ref{final_probability}.(B) The strong coupling case is characterized by $g_{strong}=g_{weak} \cdot 10^3$; in it we can still see the original beat note at ($\delta_1-\delta_2$), with additional peaks at the harmonics since for the strong coupling $\phi \sim g \tau =5$, the second order term in the expansion of $\sin(\phi)^2$ dominates, and contributes a stronger peak at the second harmonic. The simulation parameters were $g_{weak}=1$,\  $\delta_1=10^2$,\ $\delta_2-\delta_1=10^{-2}$,\ $\sigma=1$, $\Delta = 10^{-6}$. The number of measurements was $N=10^6$, and the time for each interaction was $\tau = 5 \cdot 10^{-3}$. 
\end{figure}

A few remarks are in order:
the assumption of weak coupling ($g\tau\ll1$) is not necessary. Given a general coupling strength, the probability in eq. \ref{final_probability} is generalized to:
\begin{equation}
P_{\downarrow_{\phi_{m}}}=0.5\left(1-\cos\left(\phi_{m}\right)J_{0}\left(4g\tau\frac{\sin\left(\Delta t\right)}{\Delta t}\cos\left(\frac{\delta_{1}-\delta_{2}}{2}t\right)\right)\right),
\label{strong_coupling}
\end{equation} 
where $J_{0}$ is the zeroth Bessel function.
Hence, for $\phi_m=0$ measurement, the decay time still goes as $\Delta^{-1},$ and the only difference is that this expression also contains higher harmonics of $\delta_{1}-\delta_{2}$ (see fig. \ref{variance_measurement}, \ref{simulation1}).    This expression is only valid, however, in the semi-classical regime.

The performance of this protocol depends on the measurement basis. The only terms that survive the averaging over the sample are even powers of $\phi.$
Therefore measuring in the $x$ basis($\phi_m=0$)  is optimal since the probability in this case only contains even powers of $\phi$. On the other hand this scheme does not work with $y$ basis ($\phi_m=\frac{\pi}{2}$) measurement which only yields odd powers of $\phi$. This result is seen explicitly in Eq. \ref{strong_coupling}.

\begin{figure}
\includegraphics[height=7.5cm, width=7.5cm]{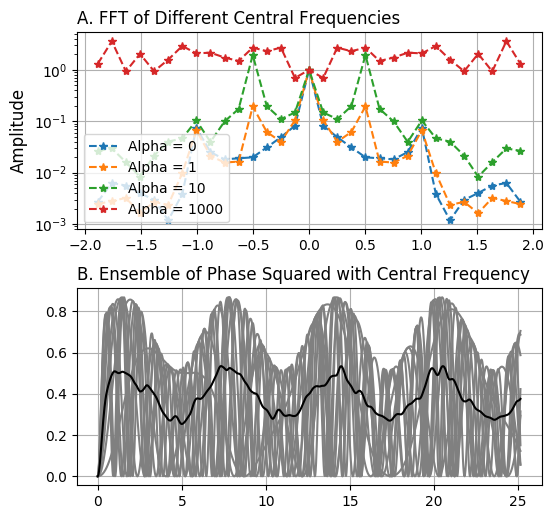}
\caption{Numerical Simulation of a strong central frequency ---}
\label{strong_central}
\raggedright
FFT of a numerical simulation of $P_{\downarrow}(t)$, in which we introduced strong central signals by simulating the phase $\phi$ using Eq. \ref{central_ferq}.
Noise was introduced by averaging $P_{\downarrow}(t)$ for $\epsilon$'s generated by a normal distribution described by $\sigma$.
(A) Without a strong central frequency; i.e. $\alpha=0$, the only notable peak is at $\delta_1 - \delta_2 = 1$. By adding a strong central signal, we get an extra peak at $\frac{\delta_1-  \delta_2}{2} = 0.5$ that is linearly proportional to the strength of the signal $\alpha$.
 The strength of the peak at $0$ is proportional to $\alpha^2$, but was normalized to $1$ for readability. For very large $\alpha$, s.t. $\phi \gg 1$, both signals are not visible.
This can be understood in the framework of Fig. \ref{variance_measurement} as follows: for small phases, $\sin(\phi)^2 \sim \phi^2$, so the probability function that measures the variance of the ensemble is now amplified by $\alpha$, as seen in (B) (an ensemble of $\phi^2$'s in grey, with their mean in black).
The variance of $\phi$ now has an element that is linear in $\alpha$, which explains the additional peak in the FFT. In the strong phase regime, the additional moments of $\phi$ are measured by $\sin(\phi)^2$ (with varying signs).
These interfering moments remove the envelope on the overall ensemble, and the signal is lost. The parameters for the simulation in (A) were $\tau = 5\cdot10^{-3}$, $g=1$, $\delta_1 = 10^2$, $\sigma = 0.1$, $\Delta = 0$ and $N=10^4$.   

\end{figure}

It is worth noting that the measurement of a chemical shift / J- coupling is more sensitive than the measurement of a single frequency because the signal of a single frequency decays on the scale of $\sigma$ and is strongly damped.
As noted earlier in the weak coupling regime the robust measurement scheme sacrifices signal. A low signal strength may for example be due to a stand-off distance of NV detectors from the target. In such a situation it would be intersting to devise a signal amplification method, one such approach closely related to optical homodyne detection will be described in the following. Indeed, an interesting effect occurs when the chemical shift appears with a background strong central frequency, $\left( \frac{\delta_1+\delta_2}{2} \right)$. This adds an interaction to the NV's Hamiltonian:

$$\footnotesize H_\text{central frequency} =  \sigma_z \left[ g \alpha \cos\left( \frac{\delta_1+ \delta_2 +2\epsilon_0 }{2}t \right) \right], $$
where $\alpha$ is the strength of central frequency. This results in the overall phase:

{\footnotesize
\begin{equation}
\phi = \left(2g\tau\right)\frac{\sin\left(\Delta t\right)}{\Delta t}\sin\left(\frac{\delta_{1}+\delta_{2}+2\epsilon_{0}}{2}t\right) \left( \cos\left(\frac{\delta_{1}-\delta_{2}}{2}t\right) + \frac{\alpha}{2}  \right).
\label{central_ferq}
\end{equation}}
When the probability is read, the chemical shift is amplified by a factor of $\alpha.$ This amplification increases the variance of the signal by a linear element in $\alpha$, so the measurement of the variance, as described in Fig \ref{variance_measurement}, should be amplified. This can be seen in the simulations shown in Fig \ref{strong_central}. The central frequency amplification is limited to the small phase regime; i.e., $\phi \sim \tau g \alpha \ll 1$, since $\phi$ does not vanish on a beat note of the chemical-shift frequency. On this beat note, different moments of $\phi$ are added together in $\sin(\phi)^2$ and interfere destructively, so that the signal is lost (see Fig \ref{strong_central} for details).

{\em Fisher Information -- }
In a given experiment, $N$ consecutive measurements of length $\tau$ will occur. The Fisher information\cite{cover2012elements} for this set of Bernoulli experiments, is:
$$
 I=\sum_{n=1}^{N}\frac{\left(\frac{\partial P(t_{n})}{\partial(\delta_{1}-\delta_{2})}\right)^{2}}{P(t_{n})(1-P(t_{n}))},
 $$
with $t_n = n \tau$. Even with a strong noise $ \sigma t \gg1$ and in the resolvable limit, i.e., $(\delta_1 - \delta_2)t\gg \pi$  this simplifies to:

\begin{eqnarray}
I=&\sum_{n=1}^{N}  \frac{4 g^{2}\tau^{4}n^{2}\left(\frac{\sin\Delta t_{n}}{\Delta t_{n}}\right)^{2}\sin^{2}\left(\frac{\delta_{1}-\delta_{2}}{2}t_{n}\right)J_{1}^{2}\left(\varphi\right)}{1-J_{0}^{2}\left(\varphi\right)}\nonumber\\&\sim\begin{cases}
\left(\frac{4}{\pi}\right)\frac{g\tau^{3}N^{3}}{3} & g\tau\gg1\\
\frac{g^{2}\tau^{4}N^{3}}{3} & g\tau\ll1
\end{cases},\label{fi}
\end{eqnarray}
with $\varphi = 2g\tau\frac{\sin\left(\Delta t\right)}{\Delta t}\cos\left(\frac{\delta_{1}-\delta_{2}}{2}t\right)$, and $J_1$ is the first Bessel function. In the weak coupling limit, we arrive at exactly the Fisher information scaling expected from a phase sensitive experiment\cite{glenn2018high,schmitt2017submillihertz, boss2017quantum,tratzmiller2019limited, rotem2019limits}. This result is of course limited by the noise measured by a single NV, apparent here in the factor $\left( \frac{\sin(\Delta t)}{\Delta t} \right)^2$.

In the case of high SNR and low coupling ($g \tau \ll 1$), $\delta_1 - \delta_2$ can be estimated by a simple Fourier transform, yielding the uncertainty derived in Eq. \ref{fi}. However this result is also valid outside this limit and can be estimated using maximum likelihood methods.
To examine behavior in the strong coupling regime (large $\phi$) we numerically simulated the process by averaging $P(t)$ (Eq. \ref{prob1_with_phi})  over a distribution of the noise $\epsilon$, with phase $\phi$ determined by Eq. \ref{Phase} (see below for details). We observed that the frequency $\delta_1-\delta_2$ was still visible in the Fourier spectrum of $P(t)$ along with its harmonics (see fig. \ref{simulation1}).

This scheme's Fisher Information scales like a phase sensitive experiment; hence, we expect longer measurements to have better sensitivity. More precisely, the variance of the measured quantity $(\delta_1 - \delta_2)$ should be proportional to $T^{-3}$ where T is the total length of the measurement.

{\em Temporal magnetic field fluctuations --} 
One setting of considerable practical relevance are temporal fluctuations in the globally applied magnetic field over long measurement cycles. Averaging over these long measurement cycles will lead to a line broadening of the individual Larmor resonances while frequency differences; e.g.,  due to chemical shifts will suffer merely the same relative broadening; in other words, if the Larmor are broadened to a width that is for example, $10^{-3}$ of the Larmor frequency; i.e., kHz, then the frequency differences are broadened to $10^{-3}$ of the chemical shift; i.e., well below 1Hz, thus remaining resolvable. 
To see this explicitly, we assume some time dependent noise $\epsilon(t)$ and define the phase accumulated due to noise $\theta(t) =  \int_0^{t} \epsilon(t') dt'$. The probability then for an $x$ basis measurement is:

{ \footnotesize
 \begin{eqnarray}
 \label{MAINTEXT_Class_Signal_Time}
P_{\downarrow_x} = \sin^2 \Bigg(g   \int_{t-\frac{\tau}{2}}^{t+\frac{\tau}{2}}  dt' \sin\left(\delta_1 t' + \theta(t')  \right) + \sin\left(\delta_2 t' +\theta(t') \right)   \Bigg) .
\end{eqnarray}
}

Averaged over the noise ensemble, it can be shown that assuming  variance of the noise $\sigma$, the amplitude of the frequency difference ($\delta_1-\delta_2$) decays as
$g^{2}\tau \sigma^{-1}$ (see SI Section-\ref{time_dependent_noise_section} for details). Given the same setup, a $y$ measurement would average a linear term over the noise, which would again result in a decay exponential in $\sigma$.

{\em Hartmann-Hahn type scheme -- } Interestingly there is an alternative approach to remove the central frequency noise,
 by decoupling the measurement probability completely from the average frequency.
This can be achieved with a Hartmann-Hahn type detection. By continuously driving the NV, we can get an effective Hamiltonian of the form:
\begin{equation}
\label{HH_detection_semiclassical}
H_{\text{eff}}=\frac{g}{2}\sum_{i=1,2}\sigma_{x}\cos\left(\delta_{i}t\right)+\sigma_{y}\sin\left(\delta_{i}t\right),
\end{equation}
see supplemental for details (SI Section-\ref*{HHDetails}). Notice that the $\sigma_x$ and $\sigma_y$ rotations have a $\pi/2$ phase with respect to the average frequency, while the beat note, $\delta_1-\delta_2$, is in sync:

\begin{eqnarray}
H_{\text{eff}}&=g\sigma_{x}\cos\left(\frac{\delta_{1}+\delta_{2}}{2}t\right)\cos\left(\frac{\delta_{1}-\delta_{2}}{2}t\right)\\&
+g\sigma_{y}\sin\left(\frac{\delta_{1}+\delta_{2}}{2}t\right)\cos\left(\frac{\delta_{1}-\delta_{2}}{2}t\right)\nonumber .
\end{eqnarray}
This will result in an effective rotation around a general axis, $\mathbf{v}(t)$, in the $\mathbf{x-y}$ plane. The rotation angle goes as the norm of $H_{\text{eff}}$ (the radius in $\mathbf{x-y}$ plane) which is completely independent from the average frequency
{\footnotesize $ H_{\text{eff}}=g\sigma_{\mathbf{v}\left(t\right)}\cos\left(\frac{\delta_{1}-\delta_{2}}{2}t\right). $}
Assuming a short interaction time $\tau \delta_i \ll 1$, we can take $\sigma_{\mathbf{v}\left(t\right)}$ to be constant. By initializing and measuring in the $\bf{z}$ axis, we only probe the $\delta_1-\delta_2$ frequency, and any noise in the $\delta_1+\delta_2$ signal will not affect our measurement. Thus, the probability for initializing and measuring $\ket{\uparrow_{z}}$ measurement is:

\begin{equation}
P_{\uparrow_{z}}=\cos^2\left(g\tau\cos\left(\frac{\delta_{1}-\delta_{2}}{2}t\right)\right),
\end{equation}
which clearly is not affected by any macroscopic noise. This scheme is still affected by microscopic noise in the same manner as described earlier (see eq. \ref{microscopic_average}). The two methods have comparable Fisher Information (see SI Section-\ref{comparison_FI}).

{\em Discussion and outlook -- } We have presented two NV - NMR Spectrometer readout techniques which are robust to field inhomogeneities.
We have analysed in detail the limit for which state of the art techniques fail to resolve the frequencies, and we have shown that the presented methods can efficiently estimate the frequency difference. We have also presented an amplification method that can improve the signal measured by the above-mentioned schemes. Our study provides strong indication that small chemical shifts and J - couplings could be estimated efficiently in the NV - NMR spectrometer setup.

{\em Acknowledgements --} The Ulm University team was supported by the ERC Synergy grant BioQ, the EU project ASTERIQS and Hyperdiamond, the BMBF via NanoSpin and DiaPol, and the DFG CRC 1279. A. R. acknowledges the support of ERC grant QRES, project No. 770929, grant agreement No 667192(Hyperdiamond), the MicroQC and ASTERIQS.

\baselineskip=12pt
\bibliography{Bib_Alex}
\bibliographystyle{apsrev4-1}

\pagebreak

{\huge Supplementary Information}
\section{The full quantum analysis} The analysis above was done in the semiclassical approximation. Here we show the full quantum analysis. We assume two sets of atoms, coupled to the NV with the same coupling constant $g$, which rotate at different frequencies $\delta_1, \delta_2$:

\begin{equation}
H = g \sigma_z \left(  \sum_i  I_x^i  + S_x^i \right)   + \sum_i \left( \frac{\delta_1}{2} I_z^i + \frac{\delta_2}{2} S_z^i\right).
\end{equation}
All atoms begin in a polarized state of $\uparrow_x$, and the NV is initialized and measured in the same scheme as mentioned above. For every specific signal the Hamiltonian is:
\begin{equation}
\label{Supp:Hamiltonian}
H =  \sum_i g \sigma_z I_x^i + H_{atom}^i,
\end{equation}
with the additional term $H_{atom}^i = \frac{\delta}{2} I_z^i$ that rotates each atom with a frequency $\delta$. We show that if we assume very little back-action, we can approximate the $I_x$ term with its expected value, $ \left< I_x\right> = \cos(\delta t)$.

In general, if at any point we can decouple the two systems $\rho_{NV}$ and $\rho_{atoms}$, we get:

\begin{eqnarray}
\label{semiclassical_approximation}
\dot{\rho}_{NV} = && -i \text{Tr}_{atoms} \left[ H, \rho_{NV} \otimes \rho_{atoms} \right]\\
=&& -i \left[ g \sigma_z, \rho_{NV} \right] \text{Tr}_{atoms} \left( \rho_{atoms} \sum_i^N I_x^i \right) \nonumber \\
=&& -i \left[ g \sigma_z \left< I_x \right>_{atoms}, \rho_{NV} \right], \nonumber
\end{eqnarray}
where the expected value of $I_x$ is the sum over all the atoms combined. This is the expression we wanted (\ref{Supp:Hamiltonian}).

Although this result is quite general, it uses the semi-classical approximation that might not be appropriate in our case, since there is some entanglement between the atoms and the NVs. Even so, it is possible to show that our approximation still holds on short time scales. Assuming all the atoms start in the same $\uparrow_x$ state, the wave function of the system is:

\begin{eqnarray}
\ket{\psi(t, \tau)} &&= \frac{\ket{\uparrow_z}^{NV} \ket{\varphi_+(t, \tau)}^{atoms} +
 \ket{\downarrow_z}^{NV} \ket{\varphi_-(t, \tau)}^{atoms}}{\sqrt{2}} \nonumber \\
 \ket{\varphi_{\pm}(t, \tau)} &&= \left(
 \frac{\cos (\frac{\delta}{2} t) e^{\pm i g \tau} \ket{\uparrow_x}-
 i \sin(\frac{\delta}{2} t) e^{\mp i g \tau} \ket{\downarrow_x} }{\sqrt{2}}\right)^N,
\end{eqnarray}
where $N$ is the number of atoms coupled to the NV. The diagonal of the density matrix $\rho_{NV}$ of the NV is constant in the $z$ basis, since the basis states are eigenstates. For $g \tau \ll 1$ and $N \gg 1$ we find:

\begin{eqnarray}
\bra{\uparrow_z} \rho_{NV} \ket{\downarrow_z} =&& \frac{1}{2}\left( \cos^2(\frac{\delta}{2} t) e^{-i 2 g \tau} + \sin^2(\frac{\delta}{2} t) e^{i 2 g \tau}\right)^N\\
=&&\frac{1}{2} \left(\cos\left(2g\tau\right)-i\sin\left(2g\tau\right)\cos\left(\delta t\right)\right)^{N}\\
\approx&& \frac{1}{2}\exp(- i 2 N g \tau \cos(\delta t ))\nonumber.
\end{eqnarray}
This is the same result we obtained from the semi-classical Hamiltonian (\ref{Supp:Hamiltonian}), which is $H = g \sigma_z \cos(\delta t)$. The diagonals are again eigenstates, and looking at the remaining elements of the density matrix for short times we find:
\begin{eqnarray}
\left\langle \uparrow_{z}\mid\rho_{NV}\mid\downarrow_{z}\right\rangle =&& \frac{1}{2} \exp\left(-i 2 \int gN\cos\left(\delta t\right)\mathrm{d}t\right)\\
\approx&&\frac{1}{2}\exp\left(-i2 N g \tau \cos\left(\delta t\right)\right) \nonumber.
\end{eqnarray}

\section{Strong noise and time dependent noise} \label{time_dependent_noise_section}In the large phase regime, due to strong constant noise or a time dependent noise, the quantum scheme still avoids the exponential decay, and only results in a $\frac{1}{\sigma}$ decay. We develop this result in what follows.

For this case, not much will change in the classical NMR setup.
The noise only appears in the $\sin^{2}\left(\frac{\delta_{1}+\delta_{2}+2\epsilon_{t}}{2}t\right)$ term, where now the noise
$\epsilon(t)$ may depend on time. This however, does not lower the sensitivity as in the time independent case.
The probability is now:
{\footnotesize
 \begin{eqnarray}
 \label{time_dependent_full}
&& P_{\downarrow_x} = \sin^2 \Bigg(g   \int_{t-\frac{\tau}{2}}^{t+\frac{\tau}{2}}  dt' \sin\left(\delta_1 t' + \theta(t')  \right) + \sin\left(\delta_2 t' +\theta(t') \right)   \Bigg)  \nonumber\\
 =&& \sin^2 \left( 2 g\int_{t-\frac{\tau}{2}}^{t+\frac{\tau}{2}} dt' \sin\left(\frac{\delta_1 t' +\delta_2 t' +2 \theta(t') }{2} \right ) \cos\left(\frac{\delta_1 -\delta_2}{2}t' \right ) \right),
\label{Class_Signal_Time}
\end{eqnarray} }
where $\theta(t) =  \int_0^{t} \epsilon(t') dt'$ where this expression has to be averaged over the microscopic and the macroscopic terms.
$ \epsilon(t')$ is a random process that can be modeled by an Ornstein Uhlenbeck process with the correlation function $ \langle \epsilon(t_1) \epsilon(t_2) \rangle = g^2 e^{ \frac{\vert  t_1 - t_2\vert}{\tau} } $ and thus $\theta(t)$ describes Brownian motion.
Thus we can assume that the previous result is still valid for times in which the random phase is much smaller than $\pi$; i.e., $\langle \theta(t)^2 \rangle \ll \pi.$ Otherwise, for large phases, there are theoretically two regimes. The correlation time can either be greater or smaller than the interaction time $\tau$, but they both converge to the same result.
The probability for $\vert {\downarrow_x} \rangle$ is again:
{\footnotesize
 \begin{equation}
 \label{time_dependent_process}
P_{\downarrow_x} = \left( 2 g \int_{t-\frac{\tau}{2}}^{t+\frac{\tau}{2}} dt' \sin\left(\frac{\delta_1 t' +\delta_2 t' +2  \theta(t')  }{2} \right ) \cos\left(\frac{\delta_1 -\delta_2}{2}t' \right )                  \right)^2,
\end{equation}}
 in the limit in which $(\delta_1 - \delta_2) \tau \ll \pi$ the $\cos$ can go out of the integral and we get:
 {\footnotesize
 \begin{equation}
P_{\downarrow_x} = 4 g^2 \left( \int_{t-\frac{\tau}{2}}^{t+\frac{\tau}{2}}  \sin\left(\frac{\delta_1 t' +\delta_2 t' +2  \theta(t')  }{2} \right ) dt'  \right)^2\cos^2\left(\frac{\delta_1 -\delta_2}{2}t \right )       .
\end{equation}}

In this case, time dependent noise behaves similarly to an average over an ensemble of time independent noises with the same variance. If we assume $\epsilon(t)$ is described by an OU process of the form $\epsilon(t+dt) = \epsilon(t) - \frac{1}{\tau_t} \epsilon(t) dt +\sigma_t dw(t),$ it can be compared to an ensemble of constant $\epsilon$'s with variance $\sigma^2 = \sigma_t^2\tau_t$. 

For such an ensemble, the region where $\left|\epsilon \tau\right| \ll 1$ dominates. Removing the $\delta_1-\delta_2$ term for the following analysis, we are interested in the average amplitude of:

 \begin{equation}
\frac{4g^{2}}{\sqrt{2\pi}\sigma}\int d\epsilon e^{-\left(\frac{\epsilon^{2}}{2\sigma^{2}}\right)}\left(\int_{t-\frac{\tau}{2}}^{t+\frac{\tau}{2}}\sin\left(\frac{\delta_{1}t'+\delta_{2}t'+2\epsilon t\prime}{2}\right)dt'\right)^{2}.
\end{equation}

\begin{figure}
\includegraphics[height=7.5cm, width=7.5cm]{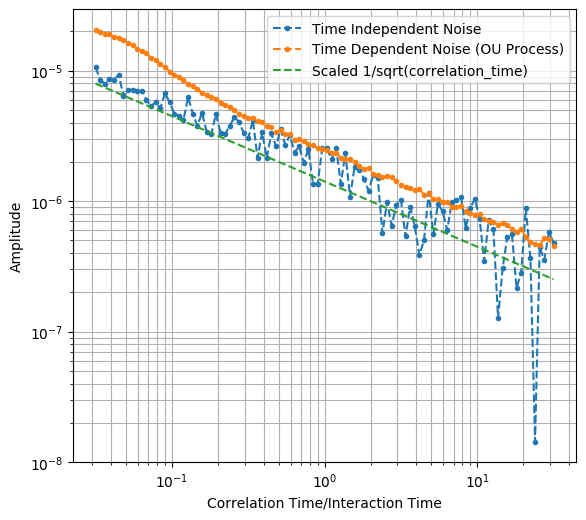}
\caption{Numerical Simulation of time dependent noise---}
\label{time_dependent_noise_figure}
\raggedright

A numerical simulation of the effect of time dependent noise on the amplitude of the $\delta_1-\delta_2$ signal (Eq. \ref{time_dependent_full}). The accumulated phase was produced by a Ornstein-Uhlenbeck process, which is described by the stochastic process : $d\epsilon = -\tau_{t}^{-1}  \epsilon dt + \sigma_t dW_t$ ($\tau_t$ is the correlation time, while $\sigma_t^2$ is the instantaneous variance). This time dependent noise process has a variance of $\sigma^2 = \sigma_t^2\tau_t$, and hence can be compared with an averaged ensemble of time independent noises with the same variance. For the time independent ensemble, $\theta(t)=\epsilon t$, with $\epsilon$ distributed with a variance $\sigma^2$ and averaged over the probability (and not the phase $\phi$). In this graph, amplitude was simulated as a function of $\tau_t / \tau \propto \sqrt{\sigma}$ (with $\tau$ being the interaction time), and a decay of $\sigma^{-1} \sim \tau_t^{-\frac{1}{2}}$ can be seen. We can see that this result is valid both when $\tau\gg\tau_t$ and $\tau \ll \tau_t$. The parameters for this simulation were $\tau=5 \cdot 10^{-3}$, $g=10^{-2}$, $N=10^4$, $\delta_1 = 10^2$, $\delta_1 -\delta_2 = 1$,$\sigma_t = 10^5  $.

\end{figure}

In the region where $\left|\epsilon \tau\right| \ll 1$, we can again assume a constant integrand to get:

\begin{equation}
\frac{4g^{2}}{\sqrt{2\pi}\sigma}\int_{\left|\epsilon\tau\right|\ll1}d\epsilon e^{-\left(\frac{\epsilon^{2}}{2\sigma^{2}}\right)}\tau^{2}\sin^{2}\left(\frac{\delta_{1}t+\delta_{2}t+2\epsilon t}{2}\right).
\end{equation}
The width of this integral is of order $\frac{1}{\tau}$, so we approximate this integral as proportional to:
\begin{equation}
\frac{4g^{2}\tau}{\sqrt{2\pi}\sigma}\left\langle\sin^{2}\left(\frac{\delta_{1}t+\delta_{2}t+2\epsilon t}{2}\right)\right\rangle_{\epsilon},
\end{equation}
and get $\sigma^{-1}$ scaling for this regime. A more detailed solution includes terms of order $\sigma^{-2}$, which in this approximation ($\sigma \tau \gg1$) should not be visible. Numerical simulations of both time dependent and time independent noises corroborate the above result (see Fig. \ref{time_dependent_noise_figure}).

\section{Numerical Simulations} To verify the above predictions, there are two types of possible experiments we would like to simulate. The first is a single NV measurement, resulting in a time series of $1,0$ depending on the result of a projective measurement. By repeating the same experiment many times and averaging the time series values, we get a time series proportional to $P(t)$, which is the probability of the projective measurement at time $t$. The second type of measurement involves a global measurement of an average luminosity of an ensemble of NV's, each either emitting a photon or not. This experiment results in a time series of a global luminosity $I(t)$, which is proportional to $P(t)$. It is worth noting that $P(t)$ is a function of the time $t$ from the beginning of the whole experiment, which is a continuous set of short experiments of length $\tau$; i.e., $\tau n = t$.

    In the simulation process, we created a time series $P(t=\tau n)$ with Eq. Main-\ref{prob1_with_phi} using $\phi(t)$ determined by the specific process we wanted to simulate and choosing $\phi_m=0$. Since $\phi$ is a function of the noise $\epsilon$, we assumed $\epsilon$ originated from a normal distribution with STD $\sigma$, and averaged the resulting $P(t)$ over the distribution. Depending on the parameters of the simulation, we either numerically integrated $P(t)$ over $\epsilon$ or generated a finite sample from the distribution and averaged the resulting $P(t)$. Integration was used whenever computationally possible.

\section{Details of the Hartmann-Hahn detection} \label{HHDetails}  Let us consider a Hartmann-Hahn type of detection; i.e., the NV sensor continuously driven by a microwave 
field (spin-lock) which is initially prepared in the $\ket{\uparrow_z}$ state, and the nuclear spins 
are rotated in the $x-y$ plane after the application of a short rf-pulse. The 
flip-flop Hamiltonian between NV and the nuclei will now aim to transfer the population to the $\ket{\downarrow_z}$ 
state. Now, crucially, let us observe over time the population of the NV center in the originally 
prepared $\ket{\uparrow_z}$. In the presence of a single Larmor frequency, the nuclei will 
create a precessing field with a fixed magnitude $\delta_1$. It is then evident that the state vector of 
the NV-center will rotate on the Blochsphere at a rate determined by the $\delta_1$ field and the overlap 
with the originally prepared $\ket{\uparrow_z}$ state will decrease irrespective of the phase as $\cos(\delta_1 t)$.
If, however, there are two nuclear species with slightly differing Larmor frequencies, the effective 
 field will be modulated at a rate that equals the difference of the two Larmor frequencies and, as a
consequence, at any moment in time the precession rate of the state vector of the NV will now depend on 
the relative phase between the Larmor precessions of the two species. Therefore, this setup is sensitive
to differences in Larmor frequencies alone and not to single Larmor frequency, at the expense of a 
reduction in signal strength that is quadratic in $\delta_1$ rather than linear.

We consider two nuclear spins that precess in the x-y-plane $(\ket{\Psi_{\text{nuc}}}=\cos\left(\delta_{1}t/2\right)\ket{+}+i\sin\left(\delta_{1}t/2\right)\ket{-})\otimes(\ket{\Psi_{\text{nuc}}}=\cos\left(\delta_{2}t/2\right)\ket{+}+i\sin\left(\delta_{2}t/2\right)\ket{-})$, acquiring phases according to their respective magnetic fields. We compare the probabilities given by an interaction with the sensing Hamiltonian:
$$ H_\text{sensing} = g \sum_i \sigma_x I_x^i, $$
with the probabilities given by the flip-flop Hamiltonian (effected by a Hartmann-Hahn scheme):
$$  H_\text{flip-flop} = \frac{g}{\sqrt{2}} \sum_i \sigma_x I_x^i + \sigma_y I_y^i,$$ 
where the index $I^i_{x,y}$ runs over the different nuclear spins, and $\sigma_{x,y}$ are the spin operators of the NV. Note that in this normalisation it is easier to compare the FI of the different methods, but it is different from the normalisation used in the semi-classical approximation (Eq. Main-\ref{HH_detection_semiclassical}). By initiating and measuring the NV in the $z$ basis, we find the probability for an $\ket{\uparrow_z}$ measurement given the Flip-Flop scheme is:

\begin{equation}
\label{hartmann_hahn_sup}
P_{\uparrow} = \frac{1}{4} \left(3+\cos (4 g \tau )- \sin ^2(2 g \tau )\cos \left(\left(\delta _1-\delta _2\right) t\right)\right), 
\end{equation}
which decouples the measurement from the frequency noise, since the probability is only sensitive to $\delta_1-\delta_2$. This is compared with an $x$ basis measurement, which doesn't decouple from the noise:
$$P_x = \frac{1}{2} + \frac{1}{\sqrt{2}}  \sin(g \tau) \cos^3(g \tau) (\cos (\delta_1 t) + \cos (\delta_2 t)).$$
For the sensing scheme, neither basis decouples from the average frequency. The probability for a same-basis measurement (i.e. $z$ basis measurement) is:
$$P_{\uparrow}=\frac{3+\cos(4g\tau)}{4}-\frac{1}{2}\sin^{2}(2g\tau)\cos\left(\delta_{1}t\right)\cos\left(\delta_{2}t\right),$$ 
and in the orthogonal $x$ basis we have:
$$P_x = \frac{1}{2} + \frac{1}{4}\sin(4g \tau) (\cos \delta_1  t+ \cos \delta_2 t),$$
neither of which decouple from the average frequency.

In our original discussion of the Hartmann-Hahn scheme, we used the exact same semi-classical approximation as in Eq. \ref{semiclassical_approximation}. The only difference is that in this case, $\left<I_y^i\right>$ also rotates. As can be seen in Eq. \ref{hartmann_hahn_sup}, treating the nuclei quantum mechanically results in the same decoupling.

Another way to obtain a signal proportional to a frequency difference and the second order in $g \tau$ is to use X initialisation, Y readout and a Hamiltonian $H_\text{eff}  = g \frac{S_x  (I_x^{(1)}+I_x^{(2)}) + S_y (I_y^{(1)}-I_y^{(2)})}{\sqrt{2}}$, but this is only possible if the frequencies are sufficiently well separated to distinguish them with a pulse sequence, and the signal $p = \frac{1}{2} - \frac{1-  \cos(4g \tau) \sin((\delta_1-\delta_2)t)}{16}$ does not give an advantage over normal X-readout.

Once this basic set up is understood, it is easier to see that there is a range of essentially equivalent alternatives
which include detection based on pulsed polarisation schemes; i.e., PulsePol \cite{Schwartz}, and \cite{machnes2010superfast,machnes2012pulsed}, where the NV state 
is initialised and measured along the same direction; {e.g., the +z-direction.} This sequence is designed to
create a flip-flop Hamiltonian analogous to the Hartmann-Hahn scheme. Furthermore, standard xy-sequences in
which initialisation and read-out are performed along the same direction will also serve this purpose.

{\it Signal for nuclear ensembles with 2 frequencies --} For many nuclei we obtain similar results. Without accounting for the measurement backaction of the NV on the nuclei 
and a sensing Hamiltonian, the readout in the Y basis signal can be approximated by \cite{IndirMeas, SchmittGS+2017}
	
	\begin{align}
	p &= \cos^2\left(\sum\limits_{m=1}^{M} g_m \cos\left(\delta_m t \right)-\frac{\pi}{4}\right)
	\\&=
	\frac{1}{2} + \frac{1}{2}\sin\left(2\sum\limits_{m=1}^{M} g_m \tau \cos\left(\delta_m t \right)\right)
	\label{signal}
	\end{align}
	
	while a readout in X-basis gives
	
	\begin{align}
	p =
	\frac{1}{2} + \frac{1}{2}\cos\left(2\sum\limits_{m=1}^{M} g_m  \tau \cos\left(\delta_m t \right)\right).
	\label{signal2}
	\end{align}
	
We assume two different frequencies leading to different phases $\delta_{1/2} t$. Again, the Y only readout contains 
 odd powers in $g \tau$, so here the individual frequencies will dominate the Fourier transform. In contrast, 
the X readout contains even powers that will also contain oscillations with the sum and the difference of these 
frequencies. As a result, the X readout will have a smaller signal, but will scale better in terms of the coupling strength.
One might expect that the weaker signal of the X-readout would hinder detection but in the absence of other
technical noise sources the Fisher Information for both readout directions is identical, as in the case of 
a single frequency \cite{SchmittGS+2017}. Simplifying to two frequencies with equal coupling, the signal at 
the nth measurement is
	
	\begin{align}
	p_n &= \cos^2\left( g \cos\left(n\tau\delta_1 \right) + g \cos\left(n\tau\delta_2 \right)-\frac{\pi}{4}\right)
\\&= \cos^2\left( 2g \cos\left(n\tau\frac{\delta_1-\delta_2}{2} \right)\cos\left(n\tau\frac{\delta_1+\delta_1}{2} \right) -\frac{\pi}{4}\right)
	\label{signal4}
	\end{align}
	
	for the Y readout and the same expression without $\pi/4$ for the X-readout. Hence, in both cases the Fisher Information on $\delta_2-\delta_1$ is
	
	\begin{align}
I=&\sum\limits _{n=1}^{N}\frac{1}{p_{n}(1-p_{n})}\left(\frac{\partial p_{n}}{\partial(\delta_{1}-\delta_{2})}\right)^{2}\\&=\tau^{2}\sum\limits _{n=1}^{N}4n^{2}g^{2}\sin^{2}\left(n\tau\frac{\delta_{1}-\delta_{2}}{2}\right)\cos^{2}\left(n\tau\frac{\delta_{1}+\delta_{2}}{2}\right)\\&\approx\tau^{2}k^{2}N^{3}/3.
	\end{align}
	
It should be noted, however, that in an experiment there may be additional noise sources. If those sources
are independent of the measurement scheme; i.e. do not scale with the signal strength, a weaker signal
can lead to a loss of SNR. In practical assessments this aspect will always need to be taken into consideration. 

\section{ Advantages of frequency differences for a single molecule and for NV ensembles} 

The Larmor frequencies of nuclear NMR signals are far larger, between $10^3$ and $10^6$ fold, 
than the frequency differences of interest that arise from chemical shifts and J-couplings. 
If the J-couplings and chemical shifts are identified as differences of Larmor frequencies,
even the smallest fluctuations in these Larmor frequencies will wash out the J-couplings and 
chemical shifts. If the relative change in frequency differences is the same as the 
absolute Larmor frequencies, schemes that are sensitive to frequency differences can
offer an advantage. This is because the X-readout makes it possible to precisely determine this desired 
frequency, whereas measuring the individual frequencies using Y-readout and inferring the 
difference from them is impossible, see Figure \ref{Fig5}. There are three principal scenarios:

{Temporal magnetic field fluctuations --} One setting of considerable practical relevance are temporal fluctuations in the globally applied
magnetic field over long measurement cycles. Averaging over these long measaurement cycles will
lead to a line broadening of the individual Larmor resonances while frequency differences; e.g.,
due to chemical shifts will suffer merely the same {\em relative} broadening; in other words, if the Larmor
are broadened to a width that is for example . $10^{-3}$ of the Larmor frequency; i.e., kHz, then the frequency
differences are broadened to $10^{-3}$ of the chemical shift; i.e., well below 1Hz, thus remaining
resolvable. 

{Spatial magnetic field fluctuations across detector ensemble --} In the first case we considered
long measurement sets obtained by a single NV where the target nuclei are subject to temporal magnetic 
field fluctuations. The measurement records obtained over time are averaged classically. This is 
equivalent to a situation in which we have a large ensemble of NV centers, e.g. close to the surface,
which observe nuclear ensembles that see a spatially homogenous magnetic field over the detection
volume of a single NV but a which may see spatial magnetic field inhomogeneities across the entire
observation region covered by all NVs. This can be the case for NVs that are implanted some distance
below a planar surface that is smaller than the lateral extent of the NV ensemble. In this situation
a phase insensitive scheme that reads out all NV centers will pick up the same signal, resulting from
the frequency difference, which can therefore be added (incoherently) thus presenting a considerable 
gain over an X-preparation-Y-readout scheme which would see strongly shifted lines upon averaging 
over many NVs, which would lead to a broadening that could easily be larger than the chemical shifts. 
	
{Diffusion --} Recall the readout of specific two individual nuclei. Assuming they 
diffuse through a magnetic field gradient that is larger than the frequency difference to be measured, 
only the X-readout allows resolution of this difference. However, this is a relatively specialised
setting as it does not apply to a larger ensemble of nuclei. In that case, frequency differences
from all over the ensemble are measured simultaneously, which leads to a strong broadening of
both Larmor frequencies and directly measured difference frequencies.

\begin{figure*}[h!]
		\centering
		\includegraphics[width=.95\linewidth]{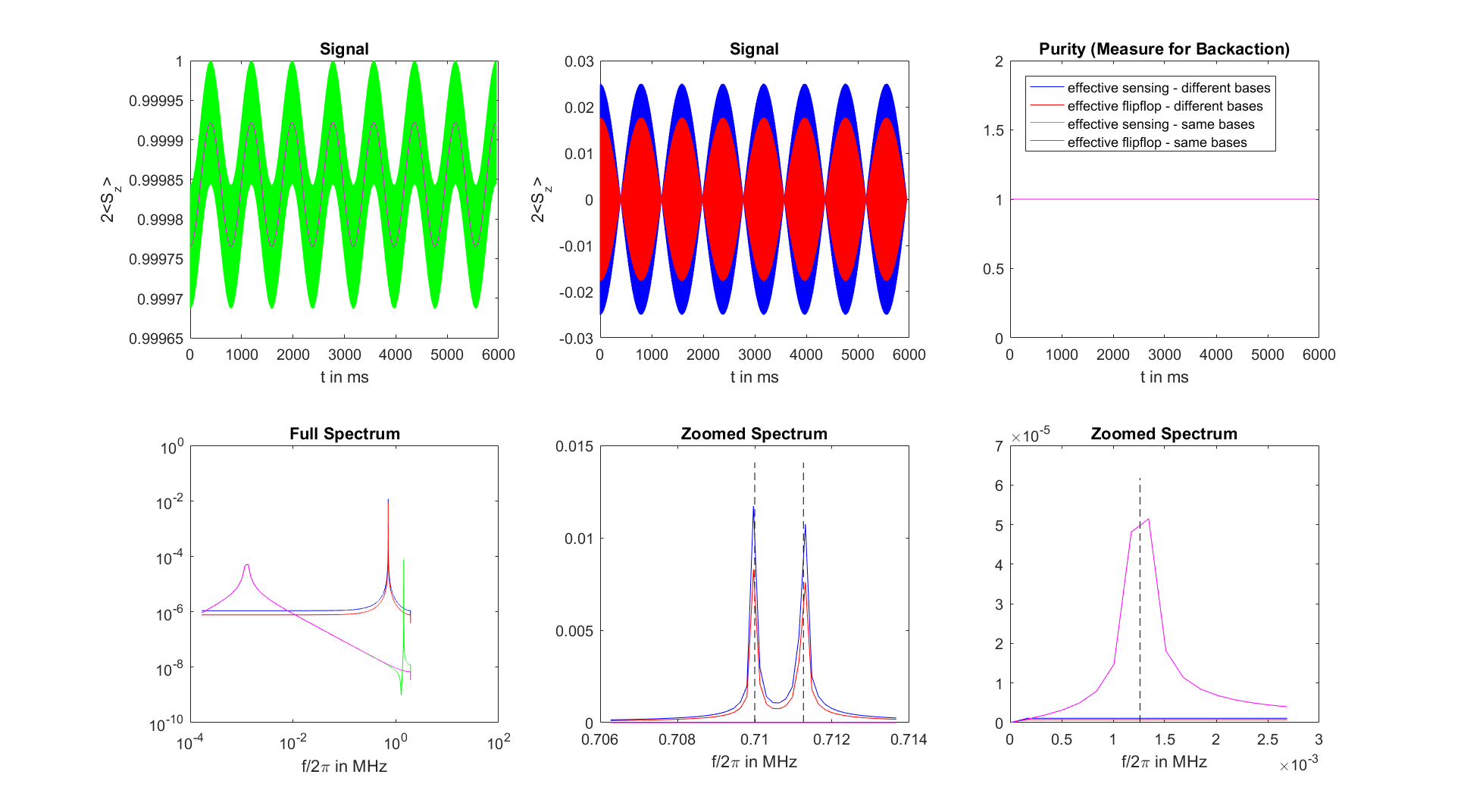}
		\caption{Simulation for 2 nuclei with $\omega_1 = (2\pi) 0.71$ MHz  and $\omega_2 = (2\pi) 0.71126$ MHz Larmor frequency, coupled with $A_x= (2\pi) 1$ kHz to the NV center without backaction with $t_s = 8\pi/\omega_1$
			a) Signal for the X-readout: the sensing Hamiltonian has additional contributions from the double frequency.
			b) Signal for the Y-readout: the sensing Hamiltonian signal is larger by a factor of 2 as expected, and the amplitude is far larger than in case of the Y-readout.
			c) Purity does not change since there is no backaction
			d) Full spectrum in a loglog-plot
			e) Zoom on the Y-readout peaks
			f) Zoom on the frequency difference peaks
		}
		\label{Fig1}
	\end{figure*}

\begin{figure*}[h!]
	\centering
	\includegraphics[width=.95\linewidth]{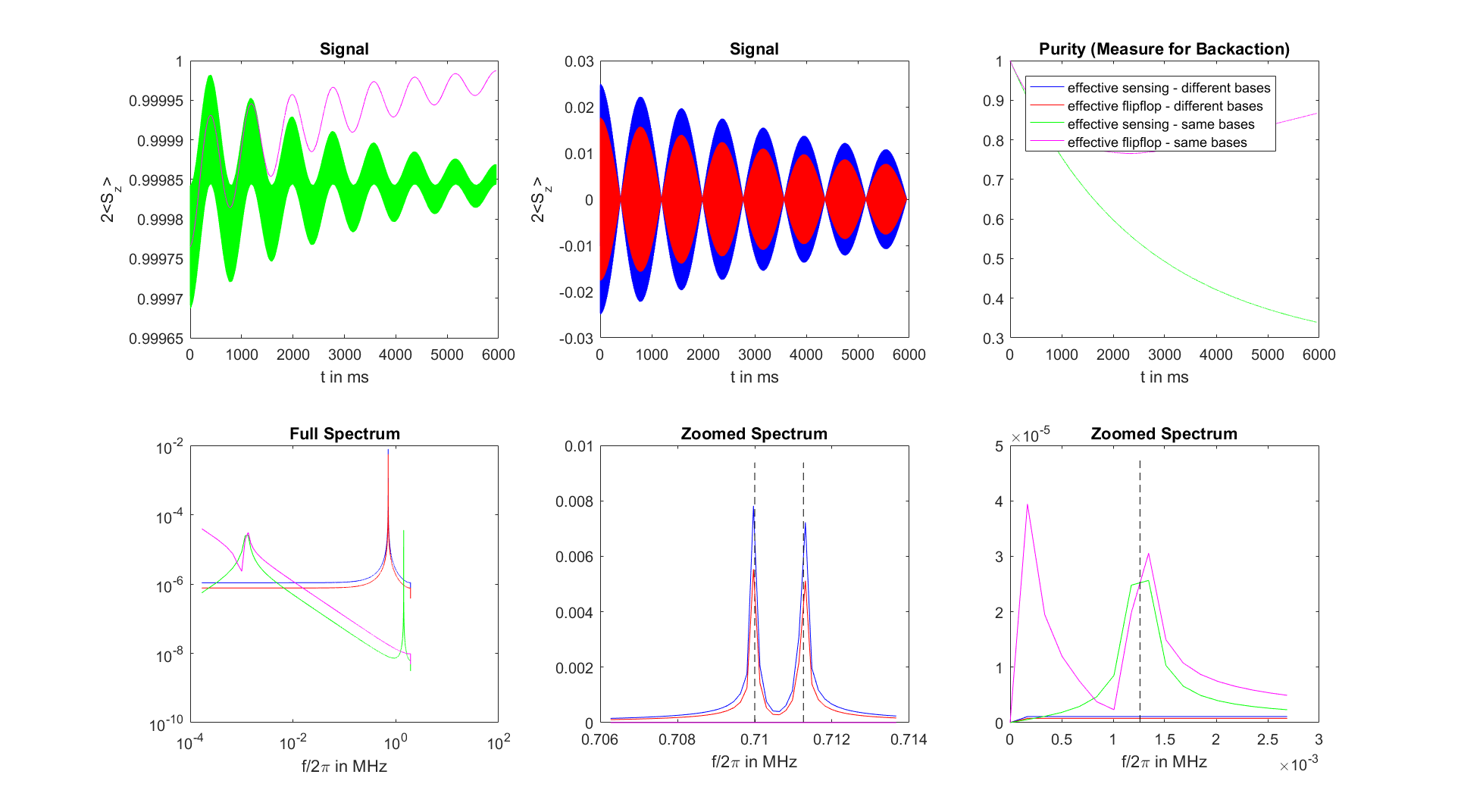}
	\caption{Parameters as in Figure \ref{Fig1}, but with backaction: for flipflop interaction the purity does not decrease, but the interaction accrues more population to $\ket{\uparrow}$ which does not contribute to the signal.
	}
	\label{Fig4}
\end{figure*}

\begin{figure*}[h!]
		\centering
		\includegraphics[width=.95\linewidth]{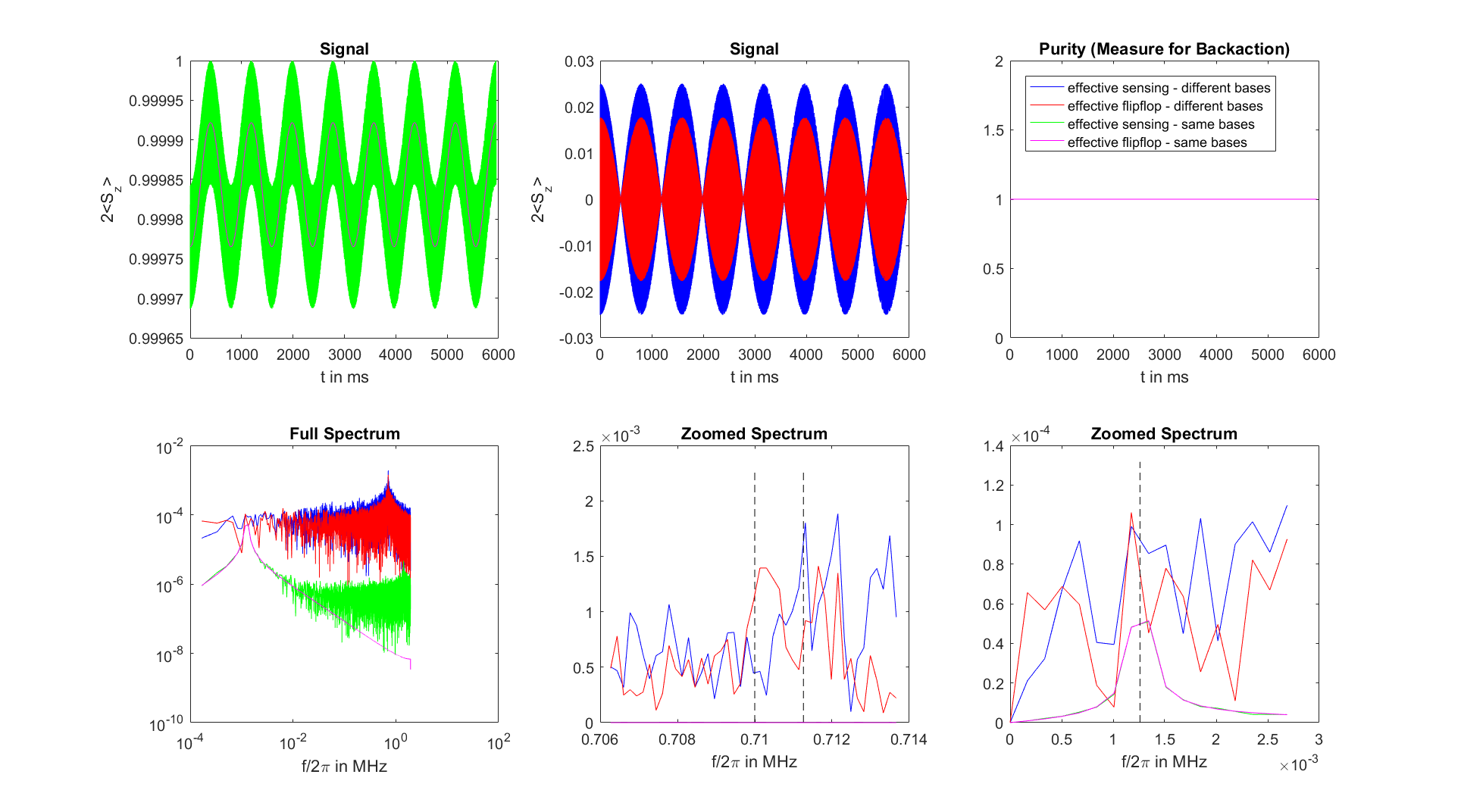}
		\caption{Parameters as in Figure \ref{Fig1}, but with temporal magnetic field fluctuations: only the frequency differences remain clear enough to distinguish small frequencies.
		}
		\label{Fig5}
	\end{figure*}

\begin{figure*}[h!]
		\centering
		\includegraphics[width=.95\linewidth]{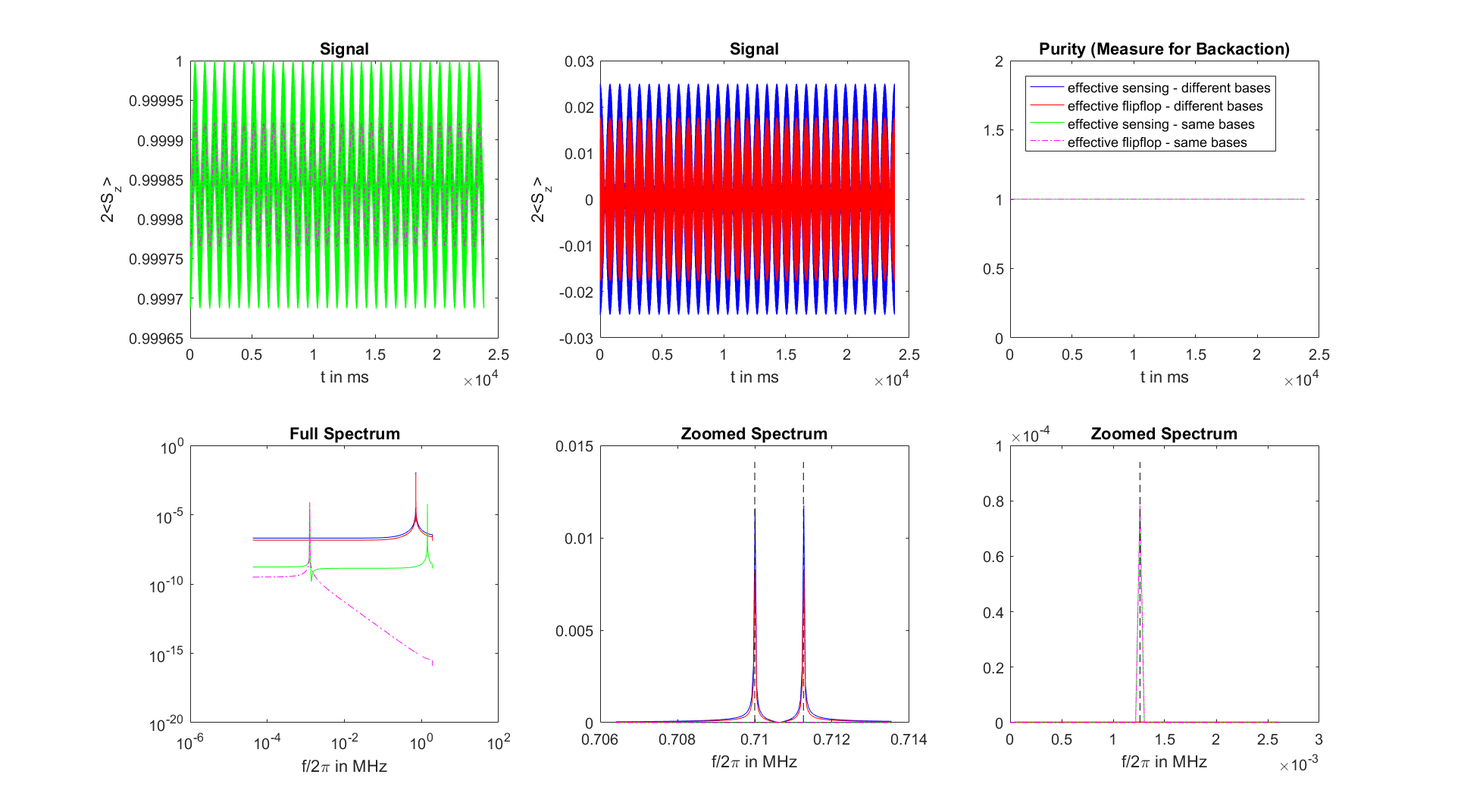}
		\caption{Parameters as in Figure \ref{Fig1}, but with longer sampling time. Now the sensing sequence in the X-readout is affected by more noise for high frequencies.
		}
		\label{Fig8}
	\end{figure*}

{\it Simulations  -}
	Several simulations are shown in the following plots. Starting from a basic example of a signal and its Fourier transformation for an X- and Y-readout in Figure \ref{Fig1}. We analyze backaction in Figure \ref{Fig4}. As a flipflop-Hamiltonian exchanges part of the state between the NV and the nuclei, the desired nuclear state is slowly replaced by a polarized state carrying no information about the desired frequencies. In contrast, backaction only decreases the state purity for a sensing Hamiltonian, see also \cite{IndirMeas}. The effects of temporal B-field fluctuations are discussed in Figure \ref{Fig5}. Figure \ref{Fig8} show different characteristics for sensing/polarization sequences for longer sampling times, in particular less noise on high frequencies for X-readout since the frequency difference alone is relevant in this case.

	Relevant effects are explained in more detail in the figure captions.
	
	\newpage
	
\section{Comparison of the Fisher Information between the sensing Hamiltonian and the Hartman Hahn}
\label{comparison_FI}
Let us compare between the FI obtained with the sensing Hamiltonian ($H_{1}$) and the HH Hamiltonian ($H_{2}$): 
\begin{eqnarray}
\begin{split}
&H_{1}=g\underset{i=1}{\overset{2}{\sum}}\cos\left(\omega_{i}t\right)\sigma_{z}\\
&H_{2}=\frac{g}{\sqrt{2}}\underset{i=1}{\overset{2}{\sum}}\cos\left(\omega_{i}t\right)\sigma_{x}+\sin\left(\omega_{i}t\right)\sigma_{y}.
\end{split}
\end{eqnarray}
Regarding $H_{1}$: Taking $\omega_{s}\left(=\frac{\omega_{1}+\omega_{2}}{2}\right)$ to be noisy, we get that the transition probability (in $\sigma_{x}$ basis) reads:
\begin{equation}
P=0.5\left(1-J_{0}\left(4g\tau \cos\left(\frac{\omega_{1}-\omega_{2}}{2}t\right)\right)\right).
\end{equation}  

The FI about $\omega_{r}=\omega_{1}-\omega_{2}$ is thus: 
\begin{eqnarray}
I\sim\begin{cases}
\frac{g^{2}\tau^{4}N^{3}}{3} & g\tau\ll1 \\
\left(\frac{4}{\pi}\right)\frac{g\tau^{3}N^{3}}{3} & g\tau\gg1
\end{cases}
\; \; ( \text{noisy } \omega_{s}) ,
\label{fi}
\end{eqnarray}

Note that for the noiseless case (where $\omega_{s}$ is not noisy) $I \sim \frac{g^{2}\tau^{4}N^{3}}{3}$ for all $g \tau.$
So interestingly for $g \tau \ll 1$ the noisy $\omega_{s}$ does not degrade the FI, for both cases we get $\frac{g^{2}\tau^{4}N^{3}}{3}.$
For $g \tau \gg 1$ we lose a factor of $\sim g \tau$ due to the noisy $\omega_{s}.$

Regarding $H_{2}$: Let us first analyze the FI for the noiseless case. 
The optimal QFI for the noiseless case reads: 
\begin{equation}
I=\underset{t}{\sum}4\tau^{2}\mu^{2}=\underset{t}{\sum}2g^{2}\tau^{2}t^{2}\sin^{2}\left(\frac{\omega_{r}}{2}t\right)\sim\frac{g^{2}\tau^{4}N^{3}}{3}.                        
\end{equation}                                                            
Initializing and measuring in the $\sigma_{z}$ basis, the transition probability and the FI read: 
\begin{equation}
p=\sin^{2}\left(\sqrt{2}g\tau\cos\left(\frac{\omega_{r}}{2}t\right)\right)\Rightarrow I=\underset{t}{\sum}2g^{2}\tau^{2}t^{2}\cos^{2}\left(\frac{\omega_{r}}{2}t\right)\sim\frac{g^{2}\tau^{4}N^{3}}{3}.                                                                                                                              
 \end{equation}
Hence this measurement scheme saturates the optimal QFI (for any $g \tau$), and is completely resilient to fluctuations of $\omega_{s}.$ 

In general, the amplitudes are not identical, hence the Hamiltonians read:                                                                                                                                                                                                                                                                                                                                                                                                                                                                                                                        
\begin{eqnarray}
\begin{split}
&H_{1}=\underset{i=1}{\overset{2}{\sum}} g_{i} \cos\left(\omega_{i}t\right)\sigma_{z}\\
&H_{2}=\frac{1}{\sqrt{2}}\underset{i=1}{\overset{2}{\sum}} g_{i} \cos\left(\omega_{i}t\right)\sigma_{x}+ g_{i} \sin\left(\omega_{i}t\right)\sigma_{y}.
\end{split}
\end{eqnarray}                                                                                                                                                                                                                                                                                                                                                                                                                                                                                                                                                                                                                                                                                                                                                                                  
Regarding $H_{1}$: Let us first analyze the FI for the noiseless case:
\begin{equation}
I\sim\underset{t}{\sum}\left(g_{1}^{2}+g_{2}^{2}\right)\tau^{2}\frac{t^{2}}{2}\approx\frac{\left(g_{1}^{2}+g_{2}^{2}\right)\tau^{4}N^{3}}{6}.
\end{equation}
This is the same as the FI about $\omega_{s},$ we get contribution from both $g_{1}, g_{2}.$
This FI however assumes knowledge of $\omega_{s}$ (it assumes that $\omega_{r}$ is the only unknown parameter), an assumption that is not valid in many cases. 
If $\omega_{s}$ is also unknown we need to consider the FI matrix of $\omega_{s},\omega_{r}$:
\begin{equation}
I=\frac{1}{6}\tau^{4}N^{3}\left(\begin{array}{cc}
g_{1}^{2}+g_{2}^{2} & 2\left(g_{1}^{2}-g_{2}^{2}\right)\\
2\left(g_{1}^{2}-g_{2}^{2}\right) & 4\left(g_{1}^{2}+g_{2}^{2}\right)
\end{array}\right).
\end{equation}
The relevant information about $\omega_{r}$ is then: 
\begin{equation}
I_{r}=\frac{1}{I_{1,1}^{-1}}=\frac{2}{3}\frac{g_{1}^{2}g_{2}^{2}}{g_{1}^{2}+g_{2}^{2}}\tau^{4}N^{3},
\end{equation}
(a simple way to understand this expression is : $I_{r}=\frac{1}{\text{var}\left(\omega_{1}\right)+\text{var}\left(\omega_{2}\right)}$).
Note that if there is a large difference between the amplitudes, $g_{1} \ll g_{2},$ $I_{r}\sim\frac{2}{3}g_{1}^{2}\tau^{4}N^{3},$ namely it goes as the minimal amplitude.  
What is the FI in the noisy case (when averaging is performed)? Note that:
\begin{equation}
p=0.5\left(1-J_{0}\left(2\tau\sqrt{g_{1}^{2}+g_{2}^{2}+2g_{1}g_{2}\cos\left(\omega_{r}t\right)}\right)\right),
\end{equation}
hence for $g_{1} \tau, g_{2} \tau \ll 1$ the FI reads: 
\begin{equation}
I=\underset{t}{\sum}\tau^{2}g_{1}^{2}g_{2}^{2}\frac{t^{2}\sin\left(\omega_{r}t\right)^{2}}{0.5\left(g_{1}^{2}+g_{2}^{2}\right)+g_{1}g_{2}\cos\left(\omega_{r}t\right)}\approx g_{ \text{min} } ^{2}\frac{\tau^{4}N^{3}}{3},
\end{equation} 
where $g_{\text{min}}=\text{min} \left( g_{1}, g_{2}  \right).$
In general the FI reads:
\begin{eqnarray}
I\sim\begin{cases}
g_{\text{min}}^{2}\frac{\tau^{4}N^{3}}{3} &   g_{1}\tau, g_{2} \tau \ll 1 \\
\frac{2}{\pi} f(c) \frac{g_{\text{min}}^{2}}{g_{\text{max}}}\frac{\tau^{3}N^{3}}{3}  &  g_{1}\tau, g_{2} \tau \gg1
\end{cases}
\end{eqnarray}
where $c=\frac{g_{\text{min}}}{g_{\text{max}}}$ and $f\left(c\right)=\frac{1}{2\pi}\overset{2\pi}{\underset{0}{\int}}\frac{\sin\left(x\right)^{2}}{\left(1+c^{2}+2c\cos\left(x\right)\right)^{1.5}}\,dx \; \left(c < 1 \right).$
$f\left(c\right)$ is increasing with $c$ and takes values between $0.5<f\left(c\right)<2.$
We get again a worse scaling for large amplitudes, and in this case we also lose due to a $\frac{g_{\text{min}}}{g_{\text{max}}}$ factor.

Regarding $H_{2}:$ If we again initialize and measure in $\sigma_{z}$ basis we get that the transition probability reads: 
\begin{equation}
p=\sin^{2}\left(\sqrt{\frac{1}{2}\left(g_{1}^{2}+g_{2}^{2}+2g_{1}g_{2}\cos\left(\omega_{r}t\right)\right)}\right).
\end{equation}
The FI is exactly like the FI of sensing for a small phase ($g_{1} \tau, g_{2} \tau \ll 1,$ just like with identical amplitudes), hence it is:\\
$I=\underset{t}{\sum}\tau^{2}g_{1}^{2}g_{2}^{2}\frac{t^{2}\sin\left(\omega_{r}t\right)^{2}}{0.5\left(g_{1}^{2}+g_{2}^{2}\right)+g_{1}g_{2}\cos\left(\omega_{r}t\right)}\approx\text{min}\left(g_{1},g_{2}\right)^{2}\frac{\tau^{4}N^{3}}{3}.$
Again, unlike $H_{1},$ this is the FI for any $g_{1},g_{2}.$ 

We remark that in this analysis we assume no back-action and thus the limit of the large phase does not apply to the quantum case in which the NMR signal is generated by a collection of molecules.

\end{document}